\documentclass[reprint, amsmath,amssymb, aps,]{revtex4-2} 	

\usepackage[dvips]{graphicx}
\usepackage{dcolumn}								
\usepackage{bm}
\usepackage{bm,amsmath}
\usepackage{braket}
\usepackage{ulem}

\usepackage{color}

\DeclareRobustCommand{\YMdel}{\bgroup\markoverwith{\textcolor[rgb]{0.1, 0.5, 0.1}{\rule[.5ex]{2pt}{0.4pt}}}\ULon}

\DeclareRobustCommand{\HHdel}{\bgroup\markoverwith{\textcolor[rgb]{1.0, 0.0, 0.0}{\rule[.5ex]{2pt}{0.4pt}}}\ULon}

\begin{document}

\preprint{APS/123-QED}

\title{Quantum annealing showing an exponentially small success probability despite a constant energy gap with polynomial energy}

\author{Hiroshi Hayasaka$^1$} 
\email{hayasaka.hiroshi@aist.go.jp}
\author{Takashi Imoto$^1$} 
\email{takashi.imoto@aist.go.jp}
\author{Yuichiro Matsuzaki$^{1,2}$}
\email{matsuzaki.yuichiro@aist.go.jp,\\ present address: ymatsuzaki872@g.chuo-u.ac.jp}
\author{Shiro Kawabata$^{1,2}$}
\email{s-kawabata@aist.go.jp,\\ present address: kawabata@hosei.ac.jp}
\affiliation{%
 $^1$Global Research and Development Center for Business by Quantum-AI Technology (G-QuAT), \\
 National Institute of Advanced Industrial Science and Technology (AIST),\\
 1-1-1, Umezono, Tsukuba, Ibaraki 305-8568, Japan\\
 $^2$NEC-AIST Quantum Technology Cooperative Research Laboratory,\\
 National Institute of Advanced Industrial Science and Technology (AIST), Tsukuba, Ibaraki 305-8568, Japan
}%
\date{\today}
\begin{abstract}   
Quantum annealing (QA) is a method for solving combinatorial optimization problems.
We can estimate the computational time for QA using what is referred to as the adiabatic condition derived from the adiabatic theorem.
The adiabatic condition consists of two parts: an energy gap and a transition matrix.
Most past  studies have focused on the relationship between the energy gap and computational time. The success probability of QA is considered  to decrease exponentially owing to the exponentially decreasing energy gap at the first-order phase-transition point. 
In  this study,  through a detailed analysis of the relationship between the energy gap, transition matrix, and computational cost during QA, we propose a general method for constructing counterintuitive models  wherein QA with a constant annealing time fails despite a constant energy gap, based on  polynomial energy.
We assume that the energy of the total Hamiltonian is at most $\Theta(L)$, where $L$ is the number of qubits.
In our formalism, we choose a known model that exhibits an exponentially small energy gap during QA, and modify the model by
adding a specific penalty term to the Hamiltonian.
In the modified model, the transition matrix in the adiabatic condition becomes exponentially large as the number of qubits increases,  while the energy gap remains constant.
Moreover, we achieve a quadratic speedup, for which the upper bound for improvement in the adiabatic condition is determined by the polynomial energy.
For concrete examples, we consider the adiabatic Grover search and
the ferromagnetic $p$-spin model.
 In these cases, with the addition of the penalty term,  although  the success probability of QA on the modified models becomes exponentially small despite a constant energy gap; we are able to achieve a success probability considerably higher than that of conventional QA. Moreover, in concrete examples, we numerically show the scaling of the computational cost is quadratically improved compared to the conventional QA. Our findings  pave the way for a better understanding of QA performance.
\end{abstract}
\pacs{Valid PACS appear here}
\maketitle

\section{\label{sec:level1}Introduction}
Quantum annealing (QA) is used to solve combinatorial optimization problems with quantum properties~\cite{Apolloni_1989, Finnila_1994, Kadowaki_1998, Farhi_2000, Farhi_2001, Arnab_2008,Albash_2018}.
The solutions to combinatorial optimization problems can be embedded in the ground state of an Ising Hamiltonian~\cite{Barahona_1982, Lucas_2014, Lechner_2015}. On the other hand, quantum fluctuation can be induced using the Hamiltonian of the transverse magnetic field. 
By gradually decreasing the transverse magnetic field while gradually increasing the Ising Hamiltonian,  we can obtain the ground state of the Ising Hamiltonian if what is known as the adiabatic condition is satisfied \cite{Kato_1950, messiah_2014, Jansen_2007, Morita_2008, Amin_2009, Kimura_2022, mackenzie2006perturbative}. 
This condition is expressed as follows: 
\begin{align}
\label{eq1}
\eta=\frac{|\braket{E_{1}(t)|\partial_{t}{\hat H_{\rm QA}(t)}|E_{0}(t)}|}{\Delta_{\rm QA}(t)^2}\ll 1,
\end{align}
where $\hat{H}_{\rm QA}(t)$, $\ket{E_{0}(t)}$, $\ket{E_{1}(t)}$, and $\Delta_{\rm QA}(t)$ are a QA Hamiltonian, the ground state, the first excited state, and the energy gap between these states, respectively.
Throughout this paper, we refer to the middle term of Eq.~\eqref{eq1} as an adiabatic-condition term.
The energy gap and computational complexity are considered to be related to each other; hence, this probable relationship has been extensively studied \cite{Seki_2012, Seki_2015, susa2022nonstoquastic, Nishimori_2017, Susa_2018, Susa_2018_2, Susa_2020}. 
Several methods for estimating and controlling the energy gap have been proposed to  improve the performance of QA \cite{Matsuzaki_2021, Imoto_2021, Mori_2022, Schiffer_2022, Imoto_2022, Kadowaki_2023}.


A phase transition can occur if there is competition between the quantum fluctuations and magnetic interactions.  
In this case, the energy gap vanishes at the point of the phase transition when the system size approaches infinity.
In statistical mechanics, the order of systems is characterized by the order parameter.
In a second-order phase transition, the order parameter is continuous, whereas in a first-order phase transition, it exhibits discontinuity at a certain time.
If a first-order phase transition occurs, it is generally supposed that the energy gap becomes exponentially small at the phase-transition point, with some exceptions \cite{Cabrera_1987, Laumann_2012, Tsuda_2013}.
When QA is used to solve hard problems (that cannot be efficiently solved using classical algorithms),  a first-order phase transition typically occurs.
For example, such a phase transition occurs when exact cover and database-search problems are solved via QA \cite{Altshuler_2010, Young_2010}.
Therefore, the adiabatic-condition term  in \eqref{eq1} is considered to increase exponentially at the first-order phase-transition point.

In the realm of quantum many-body systems, 
determining whether the energy gap is closed or open, i.e., gapless or gapped, in the thermodynamic limit is a notorious problem itself \cite{Cubitt_2015}. However, modifying the energy gap through Hamiltonian rescaling is possible, even when the energy gap is exponentially small. 
While such an energy rescaling improves the adiabatic condition term by the factor of the energy gap, this does not inherently reduce a computational cost of QA.
Using the quantum speed limit \cite{Mandelstam_1945, Vaidman_1992, Margolus_1998,   Deffner_2013, Deffner_2017}, we can evaluate the computational cost required to obtain the solution of QA with
a probability above a certain threshold \cite{Suzuki_2020, Hatomura_2021}. The cost is given as follows:
\begin{align}
Q=\int_{0}^{T}\sigma[\hat{H}_{\rm QA}(t),  \ket{\Psi(t)}]dt, \label{compcost}
\end{align}
where $\sigma[\hat{H},  \ket{\Psi}]=[\braket{\Psi|\hat{H}^2|\Psi}-\braket{\Psi|\hat{H}|\Psi}^2]^{1/2}$.
If we use an energy rescaling for QA Hamiltonian such as $e^{L}H_{QA}(t)$, the cost is not changed (see Appendix A). In other words, the cost required to obtain the solution of QA with a probability above a certain threshold cannot be improved by energy rescaling.
Hence, we do not use energy rescaling in this paper. 
Moreover, throughout our study, we assume that the QA Hamiltonian $\hat{H}_{\rm QA}(t)$ is bounded; as such, $\Theta(1)\leq \|\hat{H}_{\rm QA}(t)\|_{\rm op}\leq \Theta(L)$, where $\|\cdot \|_{\rm op}$ denotes the operator norm.

The performance of QA seems to be often determined by the energy gap. A notable exception arises in the case of perturbed Hamming weight problems, where the transition matrix (the numerator of the adiabatic-condition term) is more relevant than the energy gap, thus affecting the performance of QA \cite{Reichardt_2004, Muthukrishnan_2016}.
In this type of problem,  the transition matrix scales at most proportionally to a polynomial in  the system size. Consequently, 
the calculation can be finished within polynomial time, thus preventing catastrophic failures in QA.

In this paper, we propose a systematic method for constructing counterintuitive models wherein QA with a constant annealing time fails despite a constant energy gap.
We demonstrate that the cause of such failure is actually not the energy gap but, rather, the exponential increase in the size of the transition matrix in the adiabatic-condition term in Eq.~\eqref{eq1}. 
The key idea of our proposal is to add a penalty term $\hat{H}_{\rm pena}(t)$ to the QA Hamiltonian, which does not change the eigenstate of the Hamiltonian but changes the eigenvalue.
We analytically show that when we add such a penalty term, the transition matrix becomes exponentially large, while the energy gap remains constant. 
Furthermore, we demonstrate that we can achieve a quadratic speedup of QA by adding the penalty term to the QA Hamiltonian with polynomial energy, i.e., $\|\hat{H}_{\rm QA}(t)+\hat{H}_{\rm pena}(t)\|_{\rm op}\le \Theta (L)$, which can not be reproduced via energy rescaling. 
We also numerically perform QA on the Grover search and the ferromagnetic $p$-spin model.
The success probability of QA on these models becomes exponentially small as we increase the problem size $L$ despite an energy gap that scales as $\Theta(L^0)$ during QA. 
However, the success probability on our modified model is considerably higher than that of conventional QA. Moreover, we numerically show that the scaling of the computational cost is quadratically improved compared to the conventional QA.

The remainder of our paper is organized as follows. In Section~II, we review QA and the adiabatic Grover search.
In Section~III, we introduce the general framework for constructing a case in which the transition matrix increases exponentially.
In Section~IV, for our first example, we demonstrate an application  of our general theory on the adiabatic Grover search. In Section~V, for our second example, we show a numerical analysis on the ferromagnetic $p$-spin model. Finally, in Section~VI, we present our conclusion.

\section{Quantum annealing and adiabatic Grover search}
In this section, we review QA and adiabatic Grover search. 
\subsection{Quantum annealing}
In QA, the total Hamiltonian is defined as follows \cite{Farhi_2000, Farhi_2001}:
\begin{align}
\label{eq2}
    \hat{H}_{\rm QA}(t)=\frac{t}{T} \hat{H}_{p}+\left(1-\frac{t}{T}\right) \hat{H}_{d},
\end{align}
where $\hat{H}_{p}$ and $\hat{H}_{d}$ are the problem Hamiltonian and driver Hamiltonian, respectively.
We prepare a ground state of $\hat{H}_{d}$, and let this state evolve via the total Hamiltonian.
If an initial state at $t=0$ evolves in a sufficiently large $T$ to satisfy the adiabatic condition, we can obtain the ground state of $\hat{H}_{p}$ at $t=T$.
\subsection{Adiabatic Grover search}
Let us consider the problem of searching for a specific element in a database composed of $N$ elements. On a classical computer, we need to check, on average, half the elements to find the target element. To reduce this effort,  Grover proposed a quantum search algorithm that requires only $\sqrt{N}$ evaluations to find the target element \cite{Grover_1997}.
To search the database, we can adopt
an adiabatic algorithm  called the adiabatic Grover search
\cite{Farhi_2000, Roland_2002, Albash_2018}.  
In the adiabatic Grover search, the problem Hamiltonian is defined by
 \begin{align}
 \label{eq3}
     \hat{H}_{p}=\hat{I}-\ket{m}\bra{m}.
 \end{align}
Here, the solution to be found is denoted by
$|m\rangle $, which is represented by the computational basis ($\ket{\uparrow}$ and $\ket{\downarrow}$ ).
The number of qubits is $L$, and thus, the dimension of the Hilbert space is $2^L$. 
Meanwhile,  the driver Hamiltonian is defined by 
\begin{align}
 \label{eq4}
     \hat{H}_{d}=\hat{I}-\ket{++\cdots+}\bra{++\cdots+},
 \end{align}
where $\ket{+}$ is the eigenstate of the Pauli matrix ${\hat \sigma_{x}}$, i.e.,  $\ket{+}=(\ket{\uparrow}+\ket{\downarrow})/\sqrt{2}$.
Because the adiabatic Grover search can be 
block-diagonalized, we can analytically obtain the eigenvalues and eigenstates by diagonalizing a two-by-two matrix.
Using the desired state $\ket{m}$ and its orthogonal state $\ket{m^{\perp}}$, 
the total Hamiltonian $\hat{H}_{\rm QA}(t)$ can effectively be 
described to be
 \begin{align}
 \label{eq6}
     \hat{H}_{\rm QA}(t)=\frac{1}{2}{\hat I}-\frac{\Delta_{\rm QA}(t)}{2}{\rm cos\theta}\ \tilde{\sigma^{z}}-\frac{\Delta_{\rm QA}(t)}{2}{\rm sin\theta}\ \tilde{\sigma^{x}},
 \end{align}
 where we define
 Pauli matrices represented on the bases of $\ket{m}$ and $\ket{m^{\perp}}$ as
 $\tilde{\sigma}^{z}\equiv\ket{m}\bra{m}-\ket{m^{\perp}}\bra{m^{\perp}}$ and $\tilde{\sigma}^{x}\equiv\ket{m}\bra{m^{\perp}}+\ket{m^{\perp}}\bra{m}$.
 Here, ${\rm cos}\theta(t)$, ${\rm sin}\theta(t)$, and $\Delta_{\rm QA}(t)$ are given by
  \begin{align}
     {\rm cos}\theta(t)=\frac{1}{\Delta_{\rm QA}(t)}\left[1-2\left(1-t/T\right)(1-2^{-L}) \right],
 \end{align}
 \begin{align}
     {\rm sin}\theta(t)=\frac{2}{\Delta_{\rm QA}(t)}\left(1-t/T\right)\sqrt{2^{-L}(1-2^{-L})},
 \end{align}
 \begin{align}
 \label{gap0}
    \Delta_{\rm QA}(t)=\sqrt{\left(1-\frac{2t}{T}\right)^2+\frac{2^{-L+2}t}{T}\left(1-\frac{t}{T}\right)},
\end{align}
where we choose the branch of ${\rm Tan}^{-1}(x)$, which is given by $-\pi/2 \leq {\rm Tan}^{-1}(x) \leq \pi/2$, and define $\theta(t)={\rm Tan}^{-1}({\rm sin}\theta(t)/{\rm cos}\theta(t))$.
 The ground state $\ket{E_{0}(t)}$ and first excited state $\ket{E_{1}(t)}$ of the Hamiltonian (\ref{eq6}) are as follows:
 \begin{align}
 \label{eq10}
 \ket{E_{0}(t)}={\rm cos}\frac{\theta(t)}{2}\ket{m}+{\rm sin}\frac{\theta(t)}{2}\ket{m^{\perp}},
 \end{align}
\begin{align}
\label{eq11}
\ket{E_{1}(t)}=-{\rm sin}\frac{\theta(t)}{2}\ket{m}+{\rm cos}\frac{\theta(t)}{2}\ket{m^{\perp}}.
 \end{align}
The energies of these states are given by 
 \begin{align}
     E_{0}(t)=\frac{1}{2}(1-\Delta_{\rm QA}(t)),
 \end{align}
  \begin{align}
     E_{1}(t)=\frac{1}{2}(1+\Delta_{\rm QA}(t)).
 \end{align}
 Thus, the energy gap is $E_{1}(t)-E_{0}(t)=\Delta_{\rm QA}(t)$.
 Based on Eq.~(\ref{gap0}), the energy gap scales to $\Theta(2^{-L/2})$ at $t=T/2$.
 The numerator of the adiabatic-condition term \eqref{eq1} for the adiabatic Grover search is $\Theta(L^0)$ \cite{Roland_2002}.
 The annealing time $T$ should be scaled to $\Theta(2^L)$ to satisfy the adiabatic condition. This implies  that the time required to find the solution using the adiabatic Grover search is the same as that required for a classical search. 
  However, Roland and Cerf showed that a quadratic speedup, similar to that for Grover's algorithm, can be attained by choosing an optimal scheduling function \cite{Roland_2002}. 
 \section{General framework}                    
 In this section, we introduce  a general method for constructing models on which QA with a constant annealing time has an exponentially small success probability despite a constant energy gap.
 Let us define a system Hamiltonian as $\hat{H}(t)=\hat{H}_{\rm QA}(t) + \hat{H}_{\rm{pena}}(t)$ and the energy gap of $\hat{H}(t)$ as $\Delta(t)$,
 where 
 $H_{\rm{pena}}(t)$ denotes a Hamiltonian added to the QA Hamiltonian to improve the performance. 
Throughout our study, we assume that the system Hamiltonian and QA Hamiltonian satisfy
$\|\hat{H}(t)\|_{\rm op}\leq \Theta(L)$ and $\Theta(1)\leq \|\hat{H}_{\rm QA}(t)\|_{\rm op}\leq \Theta(L)$, respectively. Here, it is worth mentioning that a physical Hamiltonian is usually assumed to be extensive,  $\|\hat{H}(t)\|_{\rm op} = \|\hat{H}_{\rm QA}(t)\|_{\rm op} = \Theta(L)$ \cite{Kac_1963, Morita_2008, Campa_2009}. However, to consider Hamiltonian without a physical background, such as the adiabatic Grover search, we adopt assumptions, $\|\hat{H}(t)\|_{\rm op}\leq \Theta(L)$ and $\Theta(1)\leq \|\hat{H}_{\rm QA}(t)\|_{\rm op}\leq \Theta(L)$. 
We also assume that the transition matrix $|\braket{E_{1}(t)|\partial_{t}{\hat H_{\rm QA}(t)}|E_{0}(t)}|$ scales polynomially with $L$ and that the energy gap of the QA Hamiltonian $\Delta_{\rm QA}(t)$ becomes exponentially smaller as $L$ increases.
Subsequently, we show that by adding a specific penalty term to the QA Hamiltonian, we can systematically construct a model with a constant energy gap, where the transition matrix becomes exponentially large as $L$ increases.

The adiabatic-condition term can be expressed in the following form:
  \begin{align}
 \label{eq14}
\eta&=\frac{|\braket{E_{1}(t)|\partial_{t}{\hat H_{\rm QA}(t)}|E_{0}(t)}|}{\Delta_{\rm QA}(t)^2}\nonumber\\
&=\left[\frac{|\braket{E_{1}(t)|\partial_{t}{\hat H_{\rm QA}(t)}|E_{0}(t)}|}{\Delta_{\rm QA}(t)}\right]\frac{1}{\Delta_{\rm QA}(t)}.
\end{align}
The expression between the square brackets on the right-hand side of Eq.~(\ref{eq14}) yields
 \begin{align}
 \label{eq15}
     \frac{|\braket{E_{1}(t)|\partial_{t}{\hat H_{\rm QA}(t)}|E_{0}(t)}|}{\Delta_{\rm QA}(t)}=|\braket{E_{1}(t)|{d}/{dt}|E_{0}(t)}|.
 \end{align}
 Let us consider the following penalty term (to be added to the Hamiltonian $ \hat{H}_{\rm QA}(t)$):
  \begin{align}
 \label{eq16}
 \hat{H}_{\rm pena}(t)= \sum^{N-1}_{i=0}C_{i}(t)\ket{E_{i}(t)}\bra{E_{i}(t)}, 
 \end{align}
 where $N=2^L$ denotes the dimension of the Hilbert space, and $\ket{E_{i}(t)}$ is the eigenstate of $\hat{H}_{\rm QA}(t)$.
 Time-dependent coefficients $C_{i}(t)$  play a role in shifting the eigenenergies $E_{i}(t)$. 
 To maintain a constant energy gap during QA,
 we choose $C_{i}(t)$  as follows:
 \begin{align}
 \label{eq:pena_coeff}
     C_{i}(t)=E_{i}(t=0)-E_{i}(t).
 \end{align}
 Here, notably, if the operator norm of the QA Hamiltonian is bounded by $\Theta(L)$, this penalty term is also bounded by $\Theta(L)$ (see Appendix B). 
 From Eq.~(\ref{eq15}), we show that $\frac{|\braket{E_{1}(t)|\partial_{t}{\hat H_{\rm QA}(t)}|E_{0}(t)}|}{\Delta_{\rm QA}(t)}$ remains unchanged by the addition of a penalty term to the QA Hamiltonian, because the penalty term is diagonal in the energy eigenstate basis.
 This implies that
 if we add the penalty term to open an exponentially small energy gap, the transition matrix becomes exponentially large.
 For example, let us consider a QA model where the scaling of $|\braket{E_{1}(t)|\partial_{t}{\hat H_{\rm QA}(t)}|E_{0}(t)}|$ ($\Delta_{\rm QA}(t)$) is given by
 $\Theta(L^{0})$  ($\Theta(2^{-L/2})$).
 In this case, as will be shown later, by adding the penalty term (\ref{eq16}), we change the energy gap from $\Delta_{\rm QA}(t)=\Theta(2^{-L/2})$ to $\Delta(t)=\Theta(L^{0})$, while changing the transition matrix from  $|\braket{E_{1}(t)|\partial_{t}{\hat H_{\rm QA}(t)}|E_{0}(t)}|=\Theta(L^{0})$ to $|\braket{E_{1}(t)|\partial_{t}{\hat H(t)}|E_{0}(t)}|=\Theta(2^{L/2})$.
Notably, if we multiply an exponentially large factor by the QA Hamiltonian, e.g., $\hat{H}(t)=2^{L/2}\hat{H}_{\rm QA}(t)$, we obtain similar results. However, the operator norm of such a rescaled Hamiltonian is $\|\hat{H}(t)\|_{\rm op}=\Theta(2^{L/2})$, 
which violates our assumption of $\|\hat{H}(t)\|_{\rm op} \leq \Theta(L)$.

From Eqs.~(\ref{eq14}) and (\ref{eq15}), clearly, 
the improvement of the adiabatic condition due to the addition of operators for the QA Hamiltonian, such that the energy eigenstates are unchanged, is by a factor of 1/$\Delta_{\rm QA}(t)$.
To achieve optimal improvement of the adiabatic condition, where $\eta $ is minimized, we can choose the coefficient $C_{i}(t)$ such that the energy of the ground state is set to be 0, and the energies of all excited states are set to the largest eigenenergy of $\hat{H}_{\rm QA}(t)$. 
Thus, the optimal penalty term $\hat{H}^{\rm (opt)}_{\rm pena}(t)$ is given by
\begin{align}
\label{eq:pena_opt}
\hat{H}^{\rm (opt)}_{\rm pena}(t)=\sum_{i=0}^{N-1}C^{\rm (opt)}_{i}(t)\ket{E_{i}(t)}\bra{E_{i}(t)},
\end{align}
where $C^{\rm (opt)}_{i=0}(t)=-E_{0}(t)$ and $C^{\rm (opt)}_{i\neq 0}(t)=-E_{i}(t)+{\rm max}_{1\leq j \leq N-1}|E_{j}(t)|$.
In addition, we set the QA Hamiltonian to be extensive, i.e., $\|\hat{H}_{\rm QA}(t)\|_{\rm op} = \Theta(L)$ ~\cite{Morita_2008}.
 In this case, the factor $1/\Delta(t)$ is scaled to $\Theta(L^{-1})$ because $\Delta(t)={\rm max}_{1\leq i \leq N-1}|E_{i}(t)|=\|\hat{H}(t)\|_{\rm op}$ for the Hermitian operator. 
 Notably, although we considered an adiabatic condition term for the transition from the ground state to the first excited state, we could also consider transitions to other excited states. Let us define an adiabatic condition for the other excited states as
 $\eta_j=\frac{|\braket{E_{j}(t)|\partial_{t}{\hat H_{\rm QA}(t)}|E_{0}(t)}|}{(E_{j}(t)-E_{0}(t)) ^2}$ for $2\leq j\leq 2^L-1$.
 Using our modified Hamiltonian, we minimize not only $\eta$ but all $\eta_j$. 
Although Eq.~(\ref{eq:pena_opt}) is the optimal choice for the penalty term, we use Eqs.~(\ref{eq16}) and (\ref{eq:pena_coeff}) to maintain consistency hereafter.

Let us compare our penalty term with the counter-diabatic driving \cite{Demirplak_2003, Berry_2009, Guery-Odelin_2019}.  
Similar to our penalty term, the counter-diabatic term typically contains the instantaneous eigenstate of the QA Hamiltonian. 
With the counter-diabatic term,  exact adiabatic dynamics can be achieved in a shorter computational time than with traditional QA. However, if we adopt the counter-diabatic term, passing through a first-order phase-transition point during QA will require an exponentially large amount of energy (see Appendix C). By contrast, using our penalty term, we can achieve a quadratic speedup over the conventional QA with polynomial energy, i.e., $\|\hat{H}_{\rm QA}(t)+\hat{H}_{\rm pena}(t)\|_{\rm op}\le \Theta (L)$. 
 \section{Adiabatic Grover search with penalty term}
 
 For our first example, we first present an analytical investigation of the scaling of the adiabatic-condition term on the adiabatic Grover search with the penalty term (\ref{eq16}).
 Subsequently, we numerically show that the success probability of QA decreases exponentially for both our modified and conventional Hamiltonians. 
 We also show that, however, the success probability of our approach is higher than that of the conventional one. 
 \subsection{Scaling of adiabatic-condition term}   
 \begin{figure}
\includegraphics[scale=0.45,bb=580 30 0 400]{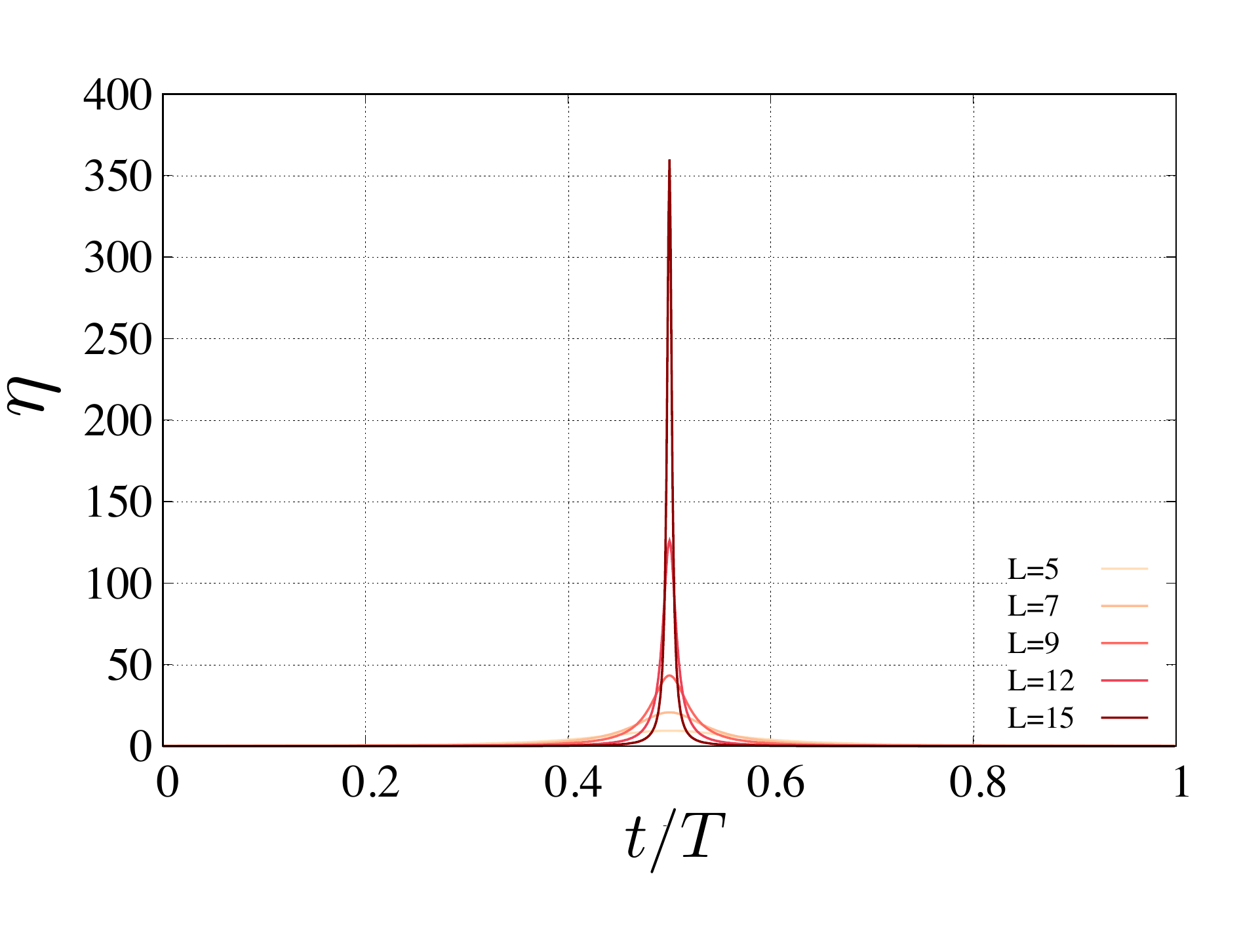} 
\caption{Plot of adiabatic-condition term $\eta$ for the adiabatic Grover search as a function of normalized time $t/T$, where $L$ is the number of qubits.} 
\label{fig1}
\end{figure}
We consider the Hamiltonian (\ref{eq6}). The penalty term is given by
 \begin{align}
 \hat{H}_{\rm pena}(t)&=-E_{0}(t)\ket{E_{0}(t)}\bra{E_{0}(t)}
 \nonumber\\&
 +\{1-E_{1}(t)\}\ket{E_{1}(t)}\bra{E_{1}(t)},
 \end{align}
 where we use Eqs.~(\ref{eq10}) and (\ref{eq11}).
 Meanwhile,  the total Hamiltonian is defined as 
 \begin{align}
 \label{eq20}
 \hat{H}(t)&=\hat{H}_{\rm QA}(t)+\hat{H}_{\rm pena}(t)\nonumber\\
  &=\frac{1}{2}{\hat I}-\frac{1}{2}{\rm cos}\theta(t)\ \tilde{\sigma}^{z}-\frac{1}{2}{\rm sin}\theta(t)\ \tilde{\sigma}^{x}.
 \end{align}
 We can easily calculate the energy gap in this model (\ref{eq20}) to be $\Delta(t)=1$.
 The energy gap in our model does not depend on the number of qubits, $L$.
Notably, the difference between Eq.~(\ref{eq20}) and  Eq.~(\ref{eq6}) is the prefactor of $\Delta_{\rm QA}(t)$ in front of ${\rm cos}\theta(t)$ and ${\rm sin}\theta(t)$. Thus, except for $\hat{I}/2$, our penalty term simply multiplies the original Hamiltonian (\ref{eq6}), by an exponentially large factor ($1/\Delta_{\rm QA}(t)$) in the subspaces of $\ket{m}$ and $\ket{m^{\perp}}$. However, $\|\hat{H}(t)\|_{\rm op}$ remains as $\Theta(1)$,  even with the penalty term.

Fig.~\ref{fig1} shows the adiabatic-condition term as a function of $t$. 
At $t=T/2$, the adiabatic-condition term becomes exponentially larger as $L$ increases.
We then analyze the scaling of the adiabatic-condition term.
From Eqs.~(\ref{eq10}), (\ref{eq11}), and (\ref{eq15}), we obtain
 \begin{align}
 \label{eq21}
     &|\braket{E_{1}(t)|d/dt|E_{0}(t)}|\nonumber\\
     &=\frac{\sqrt{(1-2^{-L})2^{-L}}}{T(1+4\left(\frac{t}{T}\right)^2(1-2^{-L})-4\frac{t}{T}(1-2^{-L}))}.
 \end{align}
 At $t=T/2$, Eq.~({\ref{eq21}}) yields 
 \begin{align}
 \label{eq22}
      |\braket{E_{1}(t)|d/dt|E_{0}(t)}|=\frac{1}{T}\sqrt{2^L-1}= \Theta(2^{L/2}).
 \end{align}
 Given that the energy gap is always constant owing to the penalty term, the scaling of the adiabatic-condition term is $\eta=$ $\Theta(2^{L/2})$ at $t=T/2$.
 This indicates that the adiabatic-condition term is improved by a factor of $\Delta_{\rm QA}^{-1}$ compared to without the penalty term.
The improvement of 1/$\Delta_{\rm QA}(t)$ is reminiscent of what is achieved using optimized scheduling on the adiabatic Grover search \cite{Roland_2002}. 
However, further improvements can be achieved by selecting an optimized schedule for the adiabatic Grover search with the penalty term (see Appendix D) \cite{Roland_2002, Albash_2018}. 
 
 Based on Eqs.~(\ref{eq15}) and (\ref{eq22}), the transition matrix becomes 
  \begin{align}
  \label{eq23}
\braket{E_{1}(t)|\partial_{t}\hat{H}(t)|E_{0}(t)}= \Theta(2^{L/2}).
 \end{align}
 Hence, as expected from our general framework, the divergence of the adiabatic-condition term in Fig.~\ref{fig1} stems from the exponentially large transition matrix (\ref{eq23}).

Quantum many-body systems often exhibit point gap closing owing to quantum phase transitions \cite{Santoro_2002, Heim_2015, Jorg_2010, Seki_2012}.
 In the adiabatic Grover search, the competition between the driver Hamiltonian and problem Hamiltonian causes a first-order quantum phase transition from the paramagnetic to ferromagnetic phase.
 The energy gap at this phase-transition point becomes exponentially small as we increase the size $L$ and vanishes at the thermodynamic limit $L\rightarrow \infty$.
 Conversely, the energy gap in our model with the penalty term
 does not close  at the thermodynamic limit, which  appears to avoid a first-order phase transition with the addition of the penalty term.
 To check whether a first-order phase transition occurs, we analyze the total magnetization as an order parameter.
 This analysis is essentially equivalent to the mean-field analysis of a $p$-spin model for $p\rightarrow \infty$ \cite{Jorg_2010}.
 Meanwhile, we directly obtain the magnetization using the ground state of our model (\ref{eq20}); this is in contrast to that done in Ref.~\cite{Jorg_2010}.
 The total magnetization of the ground state is given by
\begin{align}
\label{eq24}
    \braket{E_{0}(t)|\sum_{i}^{L}{\hat \sigma^{z}_{i}}|E_{0}(t)}={\rm cos}^{2}\frac{\theta(t)}{2}-\frac{1}{2^{L}-1}{\rm sin}^{2}\frac{\theta(t)}{2},
\end{align}
where ${\hat \sigma}^{z}$ is the $z$ component of the Pauli matrix, ${\hat \sigma}^{z}=\ket{\uparrow}\bra{\uparrow}-\ket{\downarrow}\bra{\downarrow}$.
The occurrence of a first-order phase transition can be shown as a discontinuity of magnetization at $t=T/2$ within the thermodynamic limit.
We let the thermodynamic limit be $L \rightarrow \infty$ in Eq.~(\ref{eq24}): 
\begin{align}
    \lim_{L \to \infty}\left({\rm cos}^{2}\frac{\theta(t)}{2}-\frac{1}{2^{L}-1}{\rm sin}^{2}\frac{\theta(t)}{2}\right)&=\frac{1}{2}+\frac{1}{2}\lim_{L \to \infty}{\rm cos}\theta(t) \nonumber\\
    &=\frac{1}{2}-\frac{1-2(t/T)}{2|1-2(t/T)|}.
\end{align}
Thus, we obtain
\begin{align}
  \lim_{t \to \frac{T}{2}-0}\lim_{L \to \infty}\braket{E_{0}(t)|\sum_{i}^{L}{\hat \sigma^{z}_{i}}|E_{0}(t)}=0,
\end{align}
\begin{align}
   \lim_{t \to \frac{T}{2}+0}\lim_{L \to \infty}\braket{E_{0}(t)|\sum_{i}^{L}{\hat \sigma^{z}_{i}}|E_{0}(t)}=1.
\end{align}
Therefore, in our model, a first-order phase transition occurs at $t=T/2$ even if the gap is $\Theta(L^{0})$.
Notably, there are some counter-examples in which the gap does not become exponentially small at the first-order phase-transition point \cite{Cabrera_1987, Laumann_2012, Tsuda_2013}.

\subsection{Numerical analysis}       
 \begin{figure*}[!t]
\includegraphics[scale=0.7,bb=680 30 0 500]{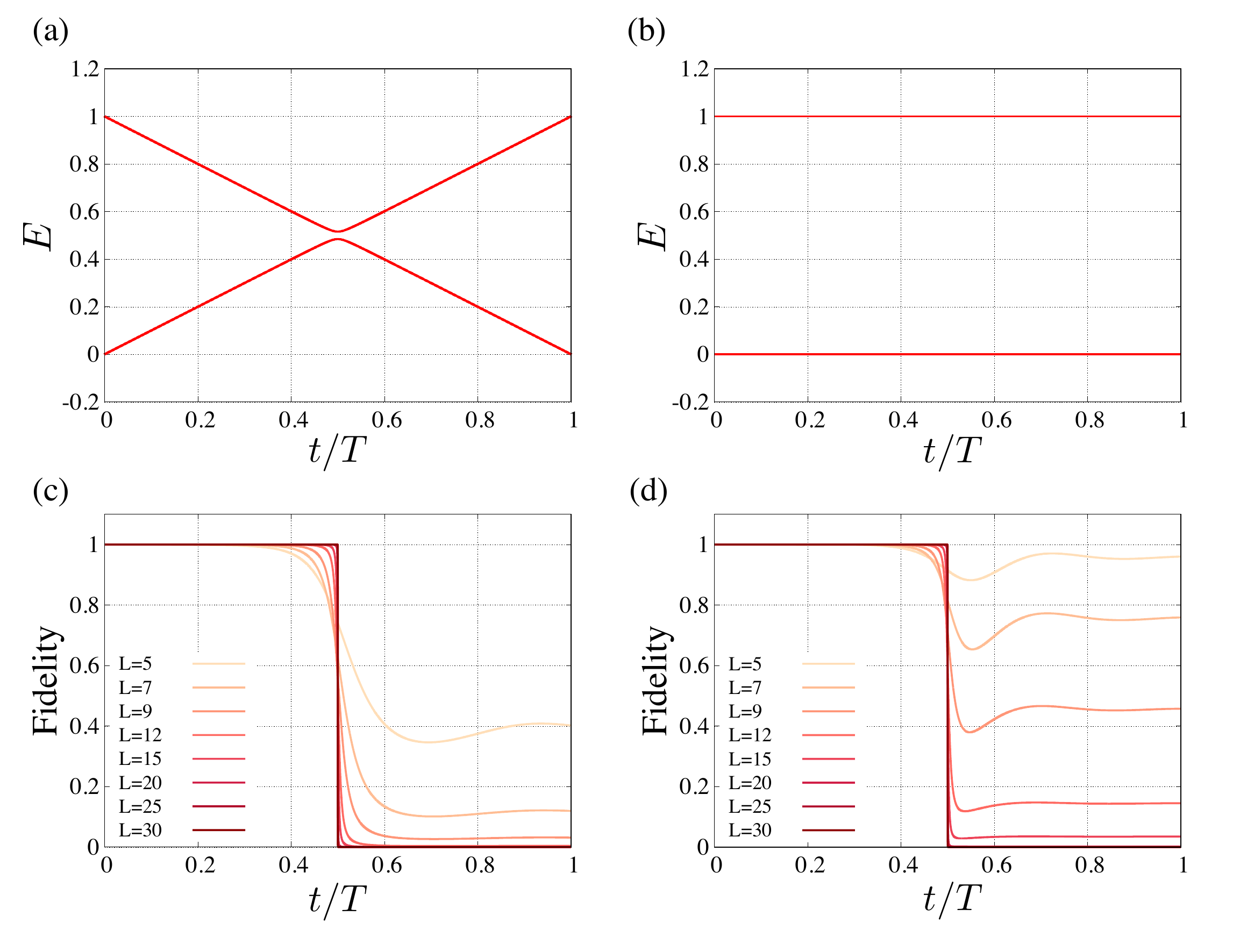}
\caption{Energy spectrum and fidelity for the adiabatic Grover search
as functions of $t/T$. Energy diagram in $L=10$ (a) without penalty term and (b) with penalty term. Fidelity (c) without penalty term and (d) with penalty term. Annealing time $T=20$.}
\label{fig2}
\end{figure*}
\begin{figure}[t]
\includegraphics[scale=0.35, bb=780 0 0 700]{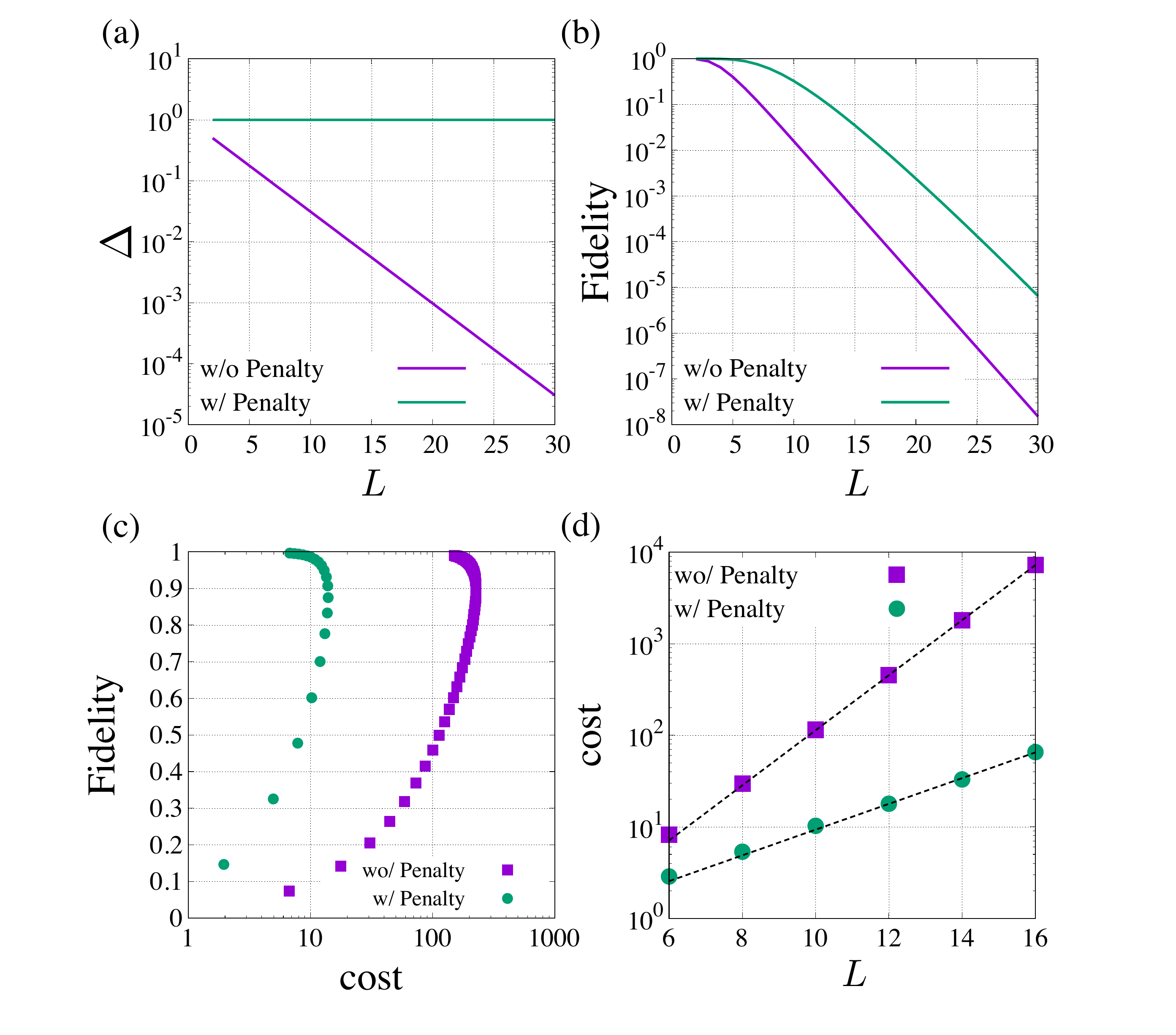}
\caption{Scaling of (a) energy gap at $t=T/2$, and (b) fidelity at $t=T$ for the adiabatic Grover search. Purple (bottom) and green (top) lines
correspond to cases without and with penalty term, respectively. In Figs. (a) and (b), annealing time is $T=20$. (c) The fidelity and corresponding
computational cost where the system size is $L=10$.
Here, we use $T=10n$ ($T=100m$) for $1\le n\le 20, n\in\mathbb{N}$ ($1\le m\le 60, m\in\mathbb{N}$)
in the case with (without) penalty term.
(d) The computational cost
to obtain a fidelity higher than 0.5.
We plot the cost against the number of qubits to identify the scaling behavior.
In Figs. (c) and (d), purple (square) and green (circle) dots
correspond to cases without and with penalty term, respectively.}
\label{fig3}
\end{figure}

 To investigate the effects of the exponential increase in size of the transition matrix of QA, we perform numerical calculations.
Fig.~{\ref{fig2}}(a) and (b) show the energy spectra without and with the penalty term, respectively. 
Without the penalty term, the energy gap becomes minimal at $t=T/2$.
By contrast, with the penalty term, the energy gap remains constant and is not affected by the number of qubits, $L$, as shown in Fig.~3(a).
We numerically solve the Schr\"{o}dinger equation and obtain the state $\ket{\Psi_{0}(t)}$.
We show the fidelity as a function of $t$ in Fig.~{\ref{fig2}}(c) and (d), where we define fidelity as $|\braket{\Psi_{0}(t)|E_{0}(t)}|^2$.
As $L$ increases, the fidelity decreases more rapidly around $t=T/2$ owing to a non-adiabatic transition to the first excited state. 

Fig.~{\ref{fig3}}(a) and (b) show the scaling of the energy gap at $t=T/2$ and the fidelity at $t=T$, respectively.
Without the penalty term, the energy gap becomes exponentially small. As mentioned in Section II, the scaling of the transition matrix is $\Theta(L^0)$ without the penalty term.
Therefore, the exponential decrease in the fidelity (bottom line in Fig.~3(b)) originates from the gap closing at $t=T/2$.
By contrast, with the penalty term, the energy gap is not affected by $L$.
Thus, the decrease in fidelity (top line in Fig.~3(b)) originates from the exponential increase in size of the transition matrix.
Notably, according to Fig.~{\ref{fig3}}(b), the fidelity for the model with the penalty term is higher than that for the model without the penalty term. 
This behavior is consistent with our general framework, demonstrating that the adiabatic-condition term with the penalty term is smaller than that without the penalty term.
Fig. 3(c) shows the computational cost defined in Eq. \eqref{compcost}. 
Here, after we choose a value of $T$, we calculate the fidelity and computational cost to be plotted. We repeat this process for several $T$, and plot these values with and without penalty term.
A non-monotonic behavior of the fidelity against the cost is observed,
and we explain the origin in the Appendix E.
From Fig. 3(c), we can estimate the required cost to achieve a certain value of fidelity.
Such a cost with the penalty term is smaller than that without the penalty term.
Fig. 3(d) shows the scaling of the cost against the number of qubits. 
By solving a time-dependent Schr\"{o}dinger equation, we can plot the fidelity against the annealing time $T$.
As we increase $T$, the fidelity also increases.
Let us consider the smallest $T$ where the corresponding fidelity
is more than or equal to 0.5. We plot the cost for such an annealing time $T$ in Fig. 3(d).
To identify the scaling behavior, 
we use a fitting function as $\alpha 2^{\beta L}$. Without the penalty term, we obtain
$\alpha=0.111$ and $\beta=0.999$. On the other hand, with the penalty term,
we obtain
$\alpha=0.365$ and $\beta=0.467$. Thus, 
the cost with the penalty term is quadratically smaller than that without the penalty term.

\section{Ferromagnetic $p$ spin model}      
\begin{figure*}[!t]
\includegraphics[scale=0.7,bb=680 30 0 500]{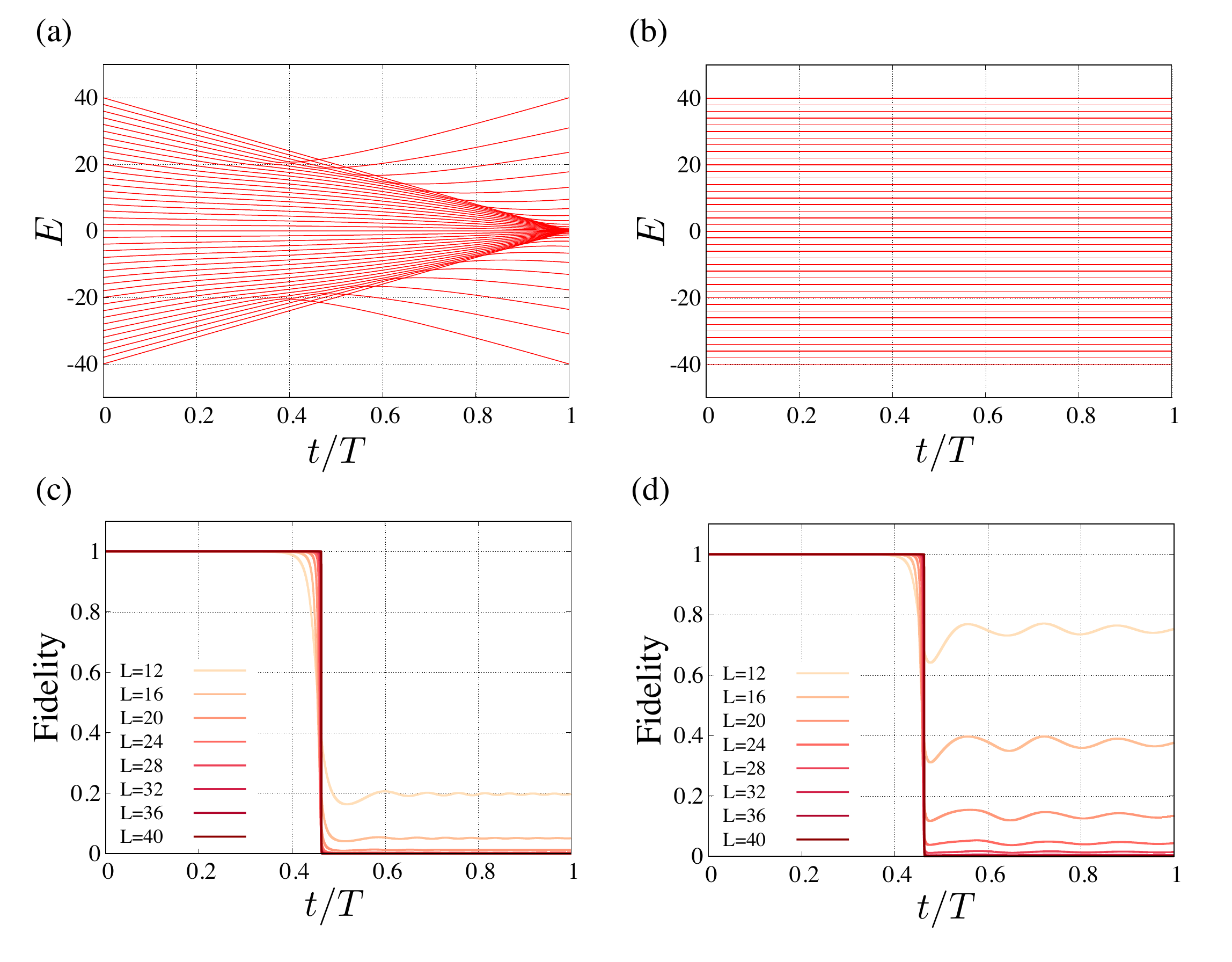}
\caption{Energy spectrum and fidelity as functions of $t/T$ for the ferromagnetic $p$-spin model. Energy diagram in $L=40$ (a) without penalty term and (b) with penalty term. Fidelity (c) without penalty term and (d) with penalty term. Annealing time $T=20$, $p=5$.} 
\label{fig4}
\end{figure*}
\begin{figure}
\includegraphics[scale=0.35, bb=780 0 0 700]{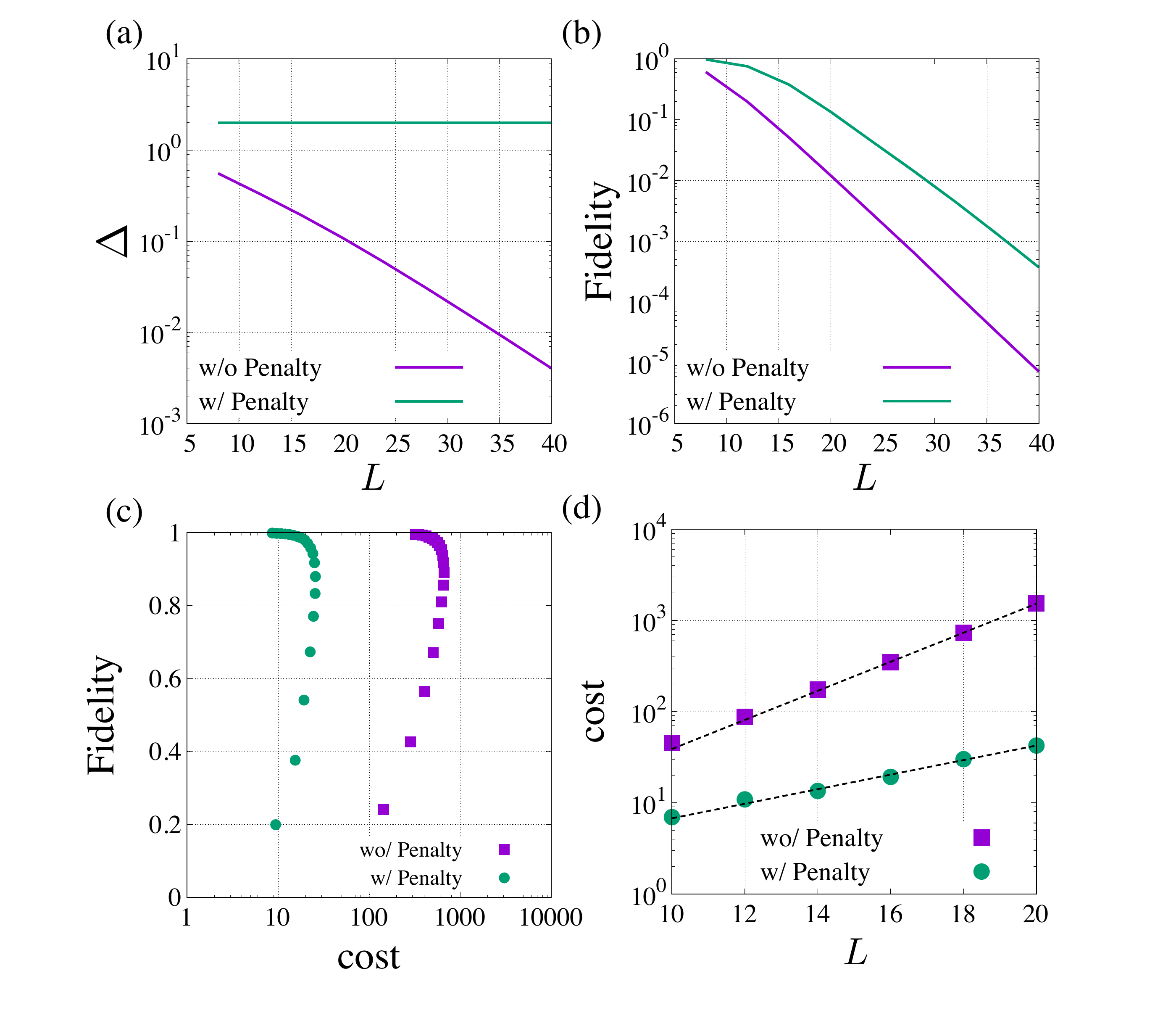}
\caption{Scaling of (a) minimum of energy gap and (b) fidelity at $t=T$ for the ferromagnetic $p$-spin model. Purple (bottom) and green (top) lines
correspond to cases without and with penalty term, respectively. In Figs. (a) and (b), we set $T=20$ and $p=5$.
(c) The fidelity and corresponding computational cost where the system size is $L=16$.
Here, we use $T=10n$ ($T=100m$) for $1\le n\le 20, n\in\mathbb{N}$ ($1\le m\le 20, m\in\mathbb{N}$) in the case with (without) penalty term.
(d) The computational cost to obtain a fidelity higher than 0.5.
We plot the cost against the number of qubits to identify the scaling behavior.
In Figs. (c) and (d), purple (square) and green (circle) dots
correspond to cases without and with penalty term, respectively.}
\label{fig5}
\end{figure}

 For our second example, we apply our theory to the ferromagnetic $p$-spin model \cite{Jorg_2010, Filippone_2011, Seki_2012, Bapst_2012}.
The problem Hamiltonian is defined as
\begin{align}
    \hat{H}_{p}=-L\left(\frac{1}{L}\sum^{L}_{i=1}\hat{\sigma}^{z}_{i}\right)^{p}.
\end{align}
We use the transverse field as the driver Hamiltonian, which is given by $\hat{H}_{d}=-\sum^{L}_{i=1}\hat{\sigma}^{x}_{i}$.
The Hamiltonian of QA is defined as follows:
\begin{align}
     \hat{H}_{\rm QA}(t)=\frac{t}{T}\hat{H}_{p}+\left(1-\frac{t}{T}\right) \hat{H}_{d}.
\end{align}
The penalty term is given by
\begin{align}
\hat{H}_{\rm pena}(t)=\sum_{i=0}^{2S}\{E_{i}({t=0})-E_{i}(t)\}\ket{E_{i}(t)}\bra{E_{i}(t)},
\end{align}
where $S$ is the maximum angular momentum and $\ket{E_{i}}$
($E_{i}$) is the eigenstate (eigenenergy) of $\hat{H}_{\rm QA}(t)$.
We numerically diagonalize $\hat{H}_{\rm QA}(t)$ each time and construct the penalty term.
The total Hamiltonian is given by
\begin{align}
    \hat{H}(t)=\hat{H}_{\rm QA}(t)+\hat{H}_{\rm pena}(t).
\end{align}
The components of the total spin operator are determined to be $\hat{S}^{z}=\sum^{L}_{i=1}\hat{\sigma}^{z}_{i}/2$ and $\hat{S}^{x}=\sum^{L}_{i=1}\hat{\sigma}^{x}_{i}/2$.
As the total spin operator $\hat{\bm{S}}$ is conserved, the initial state, belonging to the maximal spin sector, remains in this sector throughout the time evolution.
Therefore, we consider only the subspace of $2S=L$.

The ferromagnetic $p$-spin model with a transverse field for $p=2$ ($p\geq 3$) has a second-order (first-order)
phase-transition point in QA \cite{Jorg_2010, Filippone_2011, Seki_2012, Bapst_2012}.
Throughout our study, we set $p=5$.
Figs.~\ref{fig4}(a) and (b) show the energy spectra for $L=40$ without and with the penalty term, respectively. 
The minimum energy gap is at $t/T\sim 0.463$ (Fig.~\ref{fig4}(a)).
We numerically solve the Schr\"{o}dinger equation. Then, in Figs.~\ref{fig4}(c) and (d), we plot the fidelity for the model without and with the penalty term, respectively. 
In the case with the penalty term, the energy gap is always constant, as shown in Fig.~\ref{fig4}(b). 
However, at $t/T\sim 0.463$, the fidelity decreases rapidly as we
increase the number of qubits (Fig.~\ref{fig4}(d)). 
We also plot the scaling of the minimum energy gap and the fidelity at $t=T$ in Figs.~\ref{fig5}(a) and (b), respectively.
Clearly, these results are consistent with the predictions of our general framework, i.e.,
despite the constant energy gap in the case with the penalty term, the fidelity decays exponentially owing to the exponential increase in size of the transition matrix.
In addition, the fidelity of our approach is considerably higher than that of conventional QA. 
Similar to what we did in Section IV, we plot
the computational cost defined in
Eq. (2) against the fidelity
in Fig. 5(c). 
It is clearly shown that, in the presence of penalty terms, the cost is reduced compared to the case without penalty terms. We also show the scaling of the cost to achieve a fidelity greater than 0.5 in Fig. 5(d). We employ the same fitting function as in Section IV. Without penalty term, 
we obtain 
$\alpha=0.990$ and $\beta=0.530$, whereas in the presence of penalty terms, we obtain
$\alpha=1.08$ and $\beta=0.265$. Thus, 
the cost with the penalty term is quadratically smaller than that without the penalty term.
It should be noted that we ignore the cost required for diagonalization of QA Hamiltonian to construct the penalty term while we observe an improvement in the computational cost (Eq. (2)).

We compare our penalty term with the non-stoquastic Hamiltonian  in the $p$-spin model \cite{Seki_2012}. 
Although the scaling of the energy gap in the $p$-spin model can be changed from exponentially to polynomially decreasing by adding a non-stoquastic Hamiltonian, the non-stoquastic Hamiltonian does not lead to an exponential increase in the size of the transition matrix (see Appendix F). This is due to the change in not only the energy gap but also the energy eigenstate. In contrast to the non-stoquastic Hamiltonian, the modification to open the exponentially small energy gap necessarily leads to an exponential increase in the size of the transition matrix as long as we use our method. 

\section{Conclusion}
In conclusion, we propose a general method  for constructing models on which QA with a constant annealing time fails despite a constant energy gap.
In accordance with our framework, we choose a known model that exhibits an exponentially small energy gap during QA and add a penalty term to the Hamiltonian.
In the modified model, the transition matrix in the adiabatic-condition term becomes exponentially larger as the number of qubits increases, while the energy gap remains constant.
Moreover, we demonstrated that with our penalty term, the scaling of the adiabatic condition is improved from the inverse of squared to the energy gap to the inverse of the energy gap. 
We emphasize that adding the penalty term does not lead to an exponential increase in the operator norm of the Hamiltonian. Therefore, we can conclude that by using our penalty term, we can achieve quadratic speedup with polynomial energy.

Based on our framework, we investigated two models as concrete examples: the adiabatic Grover search and the ferromagnetic $p$-spin model.
For the adiabatic Grover search, we analytically showed that the transition matrix becomes exponentially large and that the magnetization exhibits discontinuity, i.e., a first-order phase transition occurs, although the energy gap is always constant in QA.
Moreover, for both the adiabatic Grover search and the ferromagnetic $p$-spin model, we numerically showed that the success probability decays exponentially owing to the exponential increase in size of the transition matrix.
In addition, we showed that the success probability with our penalty term is considerably higher than that of the conventional approach. 
Furthermore, we numerically showed the scaling of
the computational cost is quadratically improved compared to the conventional QA.

The computational speed in QA
is generally considered to be limited by the energy gap, which corresponds to the denominator of the adiabatic-condition term. Our findings prove otherwise and will lead to a deeper understanding of QA performance.

\begin{acknowledgments}
We would like to thank Yuki Susa, Ryoji Miyazaki, Yuichiro Mori, Shunsuke Kamimura, and Tadashi Kadowaki for their insightful discussions.
This study was based on the findings of the JPNP16007 project commissioned by the New Energy and Industrial Technology Development Organization (NEDO), Japan.
This work was also supported by the Leading Initiative for Excellent Young Researchers, MEXT, Japan, and JST Presto (Grant No. JPMJPR1919), Japan.
\end{acknowledgments}

\appendix
\section{Hamiltonian rescaling for computational cost}
We explain the effect of rescaled Hamiltonian given by $e^{L}\hat{H}_{\rm QA}(t)$ for the computational cost. Let us consider a time evolution by a Hamiltonian $\hat{H}_{\rm QA}(t')$ from a time $t'=0$ to $t'=t$.
We assume that $\hat{H}_{\rm QA}(t)$ is described as \eqref{eq2}.
The state is described as follows:
\begin{align}
\ket{\Psi(t)}={\cal{T}}\exp\left[-i\int_{0}^{t}\hat{H}_{\rm QA}(t')dt'\right]\ket{\Psi(0)},
\end{align}
where $\cal{T}$ is the time ordering operator. 
Now, we also consider rescaled Hamiltonian and time.
More specifically, we consider a time evolution by a Hamiltonian of $\hat{H'}_{\rm QA}(t)=e^L(\frac{t}{e^{-L}T} \hat{H}_{p}+\left(1-\frac{t}{e^{-L}T}\right) \hat{H}_{d})
$ from a time $t'=0$ to $t'=e^{-L}t$ as follows:
\begin{align}
\label{res_timeevol}
\ket{\Psi'(e^{-L}t)}={\cal{T}}\exp\left[-i\int_{0}^{e^{-L}t}\hat{H'}_{\rm QA}(t')dt'\right]\ket{\Psi(0)},
\end{align}
By changing the variable in Eq. (\ref{res_timeevol}), $t' = e^{-L} \tilde{t}$,  we obtain 
\begin{align}
&\ket{\Psi'(e^{-L}t)}\nonumber \\
&={\cal{T}}\exp\left[-i\int_{0}^{t}
(\frac{e^{-L} \tilde{t}}{e^{-L}T} \hat{H}_{p}+\left(1-\frac{e^{-L} \tilde{t}}{e^{-L}T}\right) \hat{H}_{d})
d\tilde{t}\right]\ket{\Psi(0)}\nonumber\\
&={\cal{T}}\exp\left[-i\int_{0}^{t}\hat{H}_{\rm QA}(\tilde{t})d\tilde{t}\right]\ket{\Psi(0)}\nonumber\\
&=\ket{\Psi(t)}.
\end{align}
From Eq. (A3), the time evolution of the state $\ket{\Psi'(e^{-L}t)}$ 
by $\hat{H}'_{\rm QA}(t)$
is same as that of $\ket{\Psi(t)}$ by $\hat{H}_{\rm QA}(t)$.
Thus, it is clearly shown that the Hamiltonian rescaling can shorten the annealing time from $T$ to $Te^{-L}$. 

The cost using the rescaling Hamiltonain is given by
\begin{align}
Q_{\rm res}=\int_{0}^{e^{-L}T}\sigma[\hat{H'}_{\rm QA}(t'),  \ket{\Psi'(t')}]dt'.
\end{align}
We replace time $t'$ to $t'=e^{-L}t$, then we have
\begin{align}
Q_{\rm res}&=\int_{0}^{T}\sigma[\hat{H'}_{\rm QA}(e^{-L}t),  \ket{\Psi'(e^{-L}t)}]e^{-L}dt\nonumber\\
&=\int_{0}^{T}\sigma[\hat{H}_{\rm QA}(t),  \ket{\Psi(t)}]dt\nonumber\\
&= Q.
\end{align}
Thus, the rescaling for Hamiltonian does not change the cost.
\section{Operator norm of penalty term}
In Appendix B, we show that the operator norm of the penalty term (\ref{eq16}) is bounded by $\Theta(L)$, as follows:
\begin{align}
\|\hat{H}_{\rm pena}(t)\|_{\rm op}&=\left\|\sum_{i=0}^{N-1}(E_{i}(0)-E_{i}(t))\ket{E_{i}(t)}\bra{E_{i}(t)}\right\|_{\rm op}\nonumber\\
&\leq \left\|\sum_{i=0}^{N-1}E_{i}(0)\ket{E_{i}(t)}\bra{E_{i}(t)}\right\|_{\rm op} \nonumber\\
&+\left\|\sum_{i=0}^{N-1}E_{i}(t)\ket{E_{i}(t)}\bra{E_{i}(t)}\right\|_{\rm op}\nonumber\\
&\leq \underset{0\leq i \leq N-1}{\rm max}|E_{i}(0)|+\left\|\hat{H}_{\rm QA}(t) \right\|_{\rm op}.
\end{align}
Because $\hat{H}_{\rm QA}(0)$ is Hermitian, we can show that
${\rm max}_{0\leq i \leq N-1}|E_{i}(0)|$ is equal to $\|\hat{H}_{\rm QA}(0)\|_{\rm op}$.
Thus, if the QA Hamiltonian is bounded by $\Theta(L)$ in the interval $0\leq t \leq T$, the penalty term is also bounded by $\Theta(L)$. 

\begin{figure*}[!t]
\includegraphics[scale=0.7,bb=680 30 0 300]{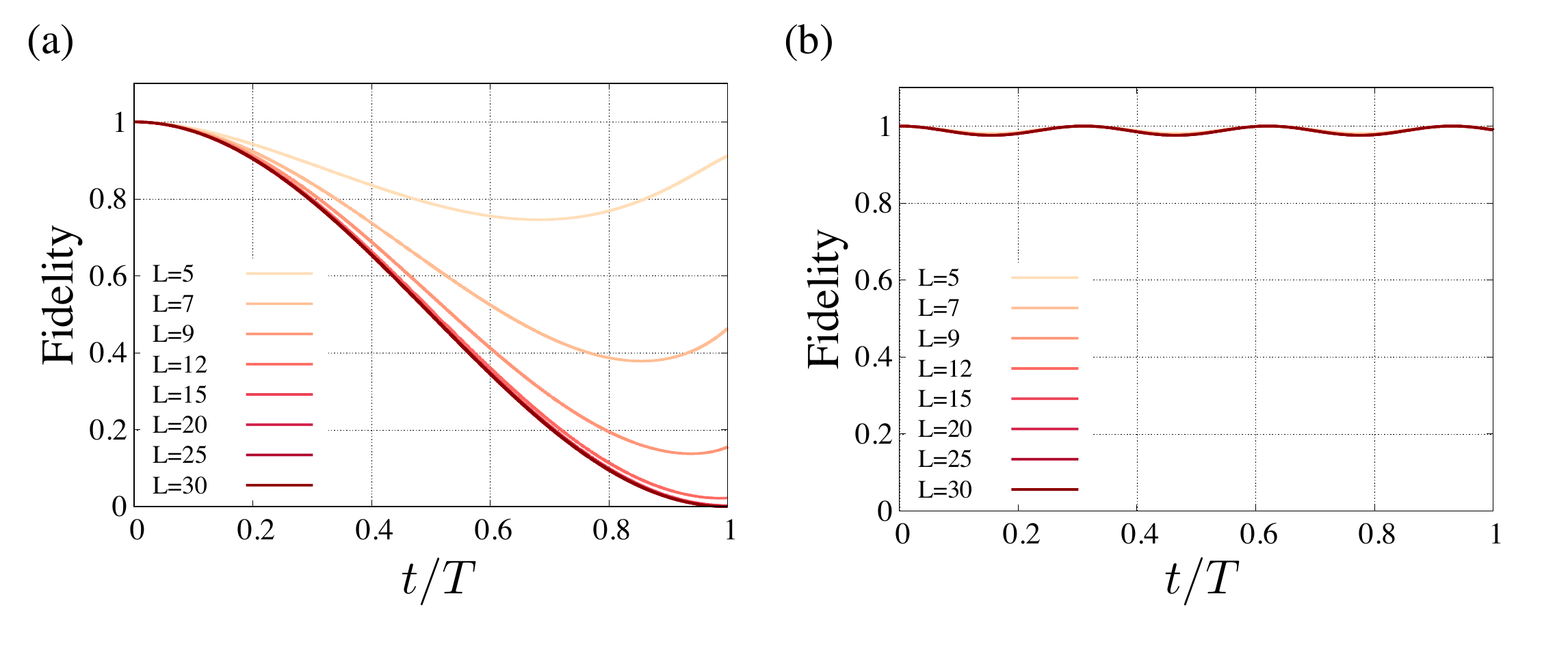}
\caption{Fidelity (a) without penalty term and (b) with penalty term as functions of $t/T$. Annealing time $T=20$.}
\label{fig6}
\end{figure*}
\begin{figure}
\includegraphics[scale=0.4,bb=580 30 0 450]{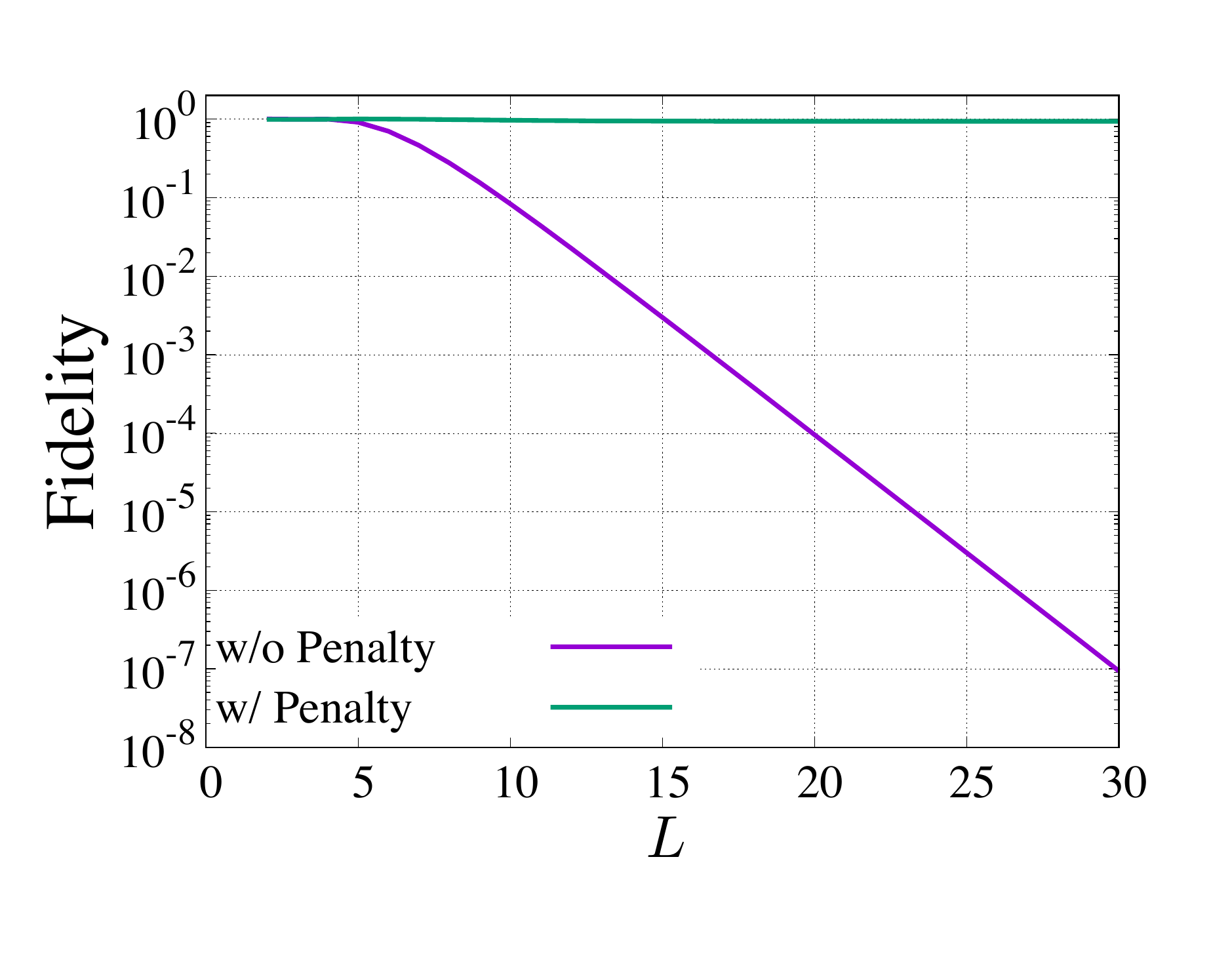}
\caption{Scaling of fidelity at $t=T$. 
Purple (bottom) and green (top) lines denote the cases without and with penalty term, respectively. Annealing time $T=20$.}
\label{fig7}
\end{figure}

\section{Operator norm of counter-diabatic term}
In Appendix C, we show that the operator norm of the counter-diabatic term becomes exponentially large at the first-order phase-transition point.
The counter-diabatic term is given by \cite{Demirplak_2003, Berry_2009, Guery-Odelin_2019}
\begin{align}
&\hat{H}_{\rm cd}(t)\nonumber\\
&=i\sum_{n}\left({\ket{\partial_{t}E_{n}(t)}\bra{E_{n}(t)}-\braket{E_{n}(t)|\partial_{t}E_{n}(t)}\ket{E_{n}(t)}\bra{E_{n}(t)}}\right).
\end{align}
The operator norm of the counter-diabatic term is given by
\begin{align}
\|\hat{H}_{\rm cd}(t)\|_{\rm op}&=\underset{{\|\ket{\Psi}\|=1}}{\rm max}\sqrt{\braket{\Psi|\hat{H}^{\dagger}_{\rm cd}(t)\hat{H}_{\rm cd}(t)|\Psi}}\nonumber\\
&=\sqrt{\sum_{n\neq m}|\braket{E_{m}(t)|\partial_{t}E_{n}(t)}|^2},
\end{align}
where $\|\cdot\|$ denotes vector norm. As shown in Section III, as the system size increases, $|\braket{E_{m}(t)|\partial_{t}E_{n}(t)}|$ increases exponentially at the first-order phase-transition point.  Thus, the operator norm of the counter-diabatic term becomes exponentially large at the first-order phase-transition point.
\section{Optimal scheduling function for the adiabatic Grover search with penalty term}

Although we chose linear scheduling in the QA Hamiltonian (Eq.~(3)) for our present study, 
we can use a more general QA Hamiltonian that uses a non-linear scheduling (but monotonously increasing) function $f(t)$ as follows:
\begin{align}
\label{eqC1}
\hat{H}_{\rm QA}=f(t)\hat{H}_{p}+(1-f(t))\hat{H}_{d}.
\end{align}
For Eq.~(\ref{eqC1}), the adiabatic condition can be modified as follows:
\begin{align}
\label{eqC2}
\eta _{\rm {gen}}\equiv \left|\frac{df(s)}{ds}\right|\frac{|\braket{E_{1}(s)|d{\hat H(s)/df}|E_{0}(s)}|}{\Delta(t)^2}\ll 1,
\end{align}
where $s$ denotes the normalized time $t/T$.
We will optimize the scheduling function to minimize $\eta _{\rm {gen}}$.
We show that through optimization, we can achieve an exponential speedup. 
From Eq.~(\ref{eqC2}), we obtain
\begin{align}
\label{eqC3}
\left|\frac{df}{ds}\right|&\leq \epsilon \frac{T\Delta(f)^2}{\left| \braket{E_{1}(t)| d\hat{H}(f)/df|E_{0}(t)}\right|}\nonumber\\
&=\epsilon  \frac{T\Delta(f)}{|\braket{E_{1}(t)|d/df|E_{0}(t)}|}  \ \ \ (\epsilon \ll 1).
\end{align}

The scheduling function is optimized if equality holds in Eq.~(\ref{eqC3}). We have
\begin{align}
\label{eqC4}
\frac{df}{ds}=c\epsilon T (|\braket{E_{1}(t)|d/df|E_{0}(t)}|)^{-1}.
\end{align}
Here, $\Delta(t)$ is now $\Theta(L^0)$; and thus, we set it to $c$. Substituting Eq.~(\ref{eq21}) into Eq.~(\ref{eqC4}), we obtain
\begin{align}
\label{eqC5}
\frac{ds}{df}&=\frac{1}{c\epsilon T} \frac{\sqrt{(1-x)\ x}}{1+4f^2(1-x)-4f(1-x)}\nonumber\\
&=\frac{1}{c\epsilon T}\frac{\sqrt{(1-x)\ x}}{1-4(1-x)f(1-f)},
\end{align}
for which we define $x=2^{-L}$. We solve this differential equation at $s(f=0)=0$ and obtain
\begin{align}
\label{eqC6}
    s=\frac{1}{2c\epsilon T} \left[{\rm Tan}^{-1}\sqrt{\frac{1-x}{x}} +{\rm Tan}^{-1}\left((2f-1)\sqrt{\frac{1-x}{x}}\right)\right].
\end{align}
By solving this equation in reverse, we obtain the optimized scheduling function:
\begin{align}
\label{eqC7}
    f(s)=\frac{1}{2}+\frac{1}{2}\sqrt{\frac{x}{1-x}}{\rm tan}\left[(2s-1){\rm Tan}^{-1}\sqrt{\frac{1-x}{x}}\right].
\end{align}
Considering the limit $x\rightarrow 0$ at $f(s=1)=1$, 
we have 
\begin{align}
\label{eqC8}
    T=\frac{1}{c\epsilon}{\rm Tan}^{-1}\sqrt{\frac{1-x}{x}}\rightarrow \frac{\pi}{2c\epsilon}.
\end{align}
Thus, the annealing time $T$ does not depend on $L$.
This finding can be intuitively understood as follows. 
While the penalty term provides a factor of $2^{L/2}$ to the adiabatic condition, optimization of the scheduling function also provides a factor of $2^{L/2}$ to it. Thus, in total, we obtain an improvement by a factor of $2^{L}$. 

To validate our analytical results, we performed numerical simulations. 
  Figs.~\ref{fig6} and \ref{fig7} show the time dependence of the fidelity and scaling of the fidelity at $t=T$, respectively. The annealing time was set to $T=20$.
As expected from Eq.~(\ref{eqC8}), when the penalty term and optimized scheduling are used, the fidelity at $t=T$ becomes $\Theta(L^0)$.

\section{Cost vs fidelity in two-level system}
To explain a non-monotonic behavior of the fidelity as a function of the cost in Fig. 3(c), 
 we consider a simple case that the system is in a superposition of the ground and first excited states.
 Let us consider a state as follows:
\begin{align}
\ket{\Psi(t)}=\frac{1}{\sqrt{1+|\epsilon|^2}}\ket{E_{0}(t)}+\frac{\epsilon}{\sqrt{1+|\epsilon|^2}}\ket{E_{1}(t)}.
\end{align}
Using this state, the variance becomes
\begin{align}
&\braket{\Psi(t)|\hat{H}^2(t)|\Psi(t)}-\braket{\Psi(t)|\hat{H}(t)|\Psi(t)}^2\nonumber\\
&=\frac{1}{1+|\epsilon|^2}(E_{0}^2(t)+|\epsilon|^2E_{1}^2(t))\nonumber\\
&-\left(\frac{1}{1+|\epsilon|^2}\right)^2(E_{0}^2(t)+2|\epsilon|^2E_{0}(t)E_{1}(t)+|\epsilon|^4E_{1}^2(t)).
\end{align}
For simplicity, let us consider the QA Hamiltonian with penalty term. In this case, we can set $E_{0}(t)=0$ and $E_{1}(t)=1$. We obtain the standard deviation as follows:
\begin{align}
\sigma=\frac{|\epsilon|}{1+|\epsilon|^2}.
\end{align}
Since the fidelity is given by $F=|\braket{E_{0}(t)|\Psi(t)}|^2=1/(1+|\epsilon|^2)$, the standard deviation becomes
\begin{align}
\sigma=\sqrt{F(1-F)}.
\end{align}
To proceed with further calculations, 
we assume that an asymptotic form of the fidelity $F$ is given by $F=1+a_1/T+a_2/T^2$.
Also, we assume that the cost is given by $T\sqrt{F(1-F)}$ for simplicity. Then we have
\begin{align}
Q=\sqrt{-a_{1}^2-a_{2}-a_{1}T}+{\cal O}(T^{-1/2}).
\end{align}
To determine parameters $a_{1}$ and $a_{2}$, we use $F=1+a_1/T+a_2/T^2$ as a fitting function to reproduce the numerical results of the fidelity against the annealing time in the adiabatic Grover search with penalty term (Fig. 8).
We obtain $a_{1}=4.31$ and $a_{2}=-976$. 
Since $a_1$ is positive, the cost becomes a decreasing function of $T$ in the region with large $T$.

\begin{figure}[!t]
\includegraphics[scale=0.4,bb=580 30 0 450]{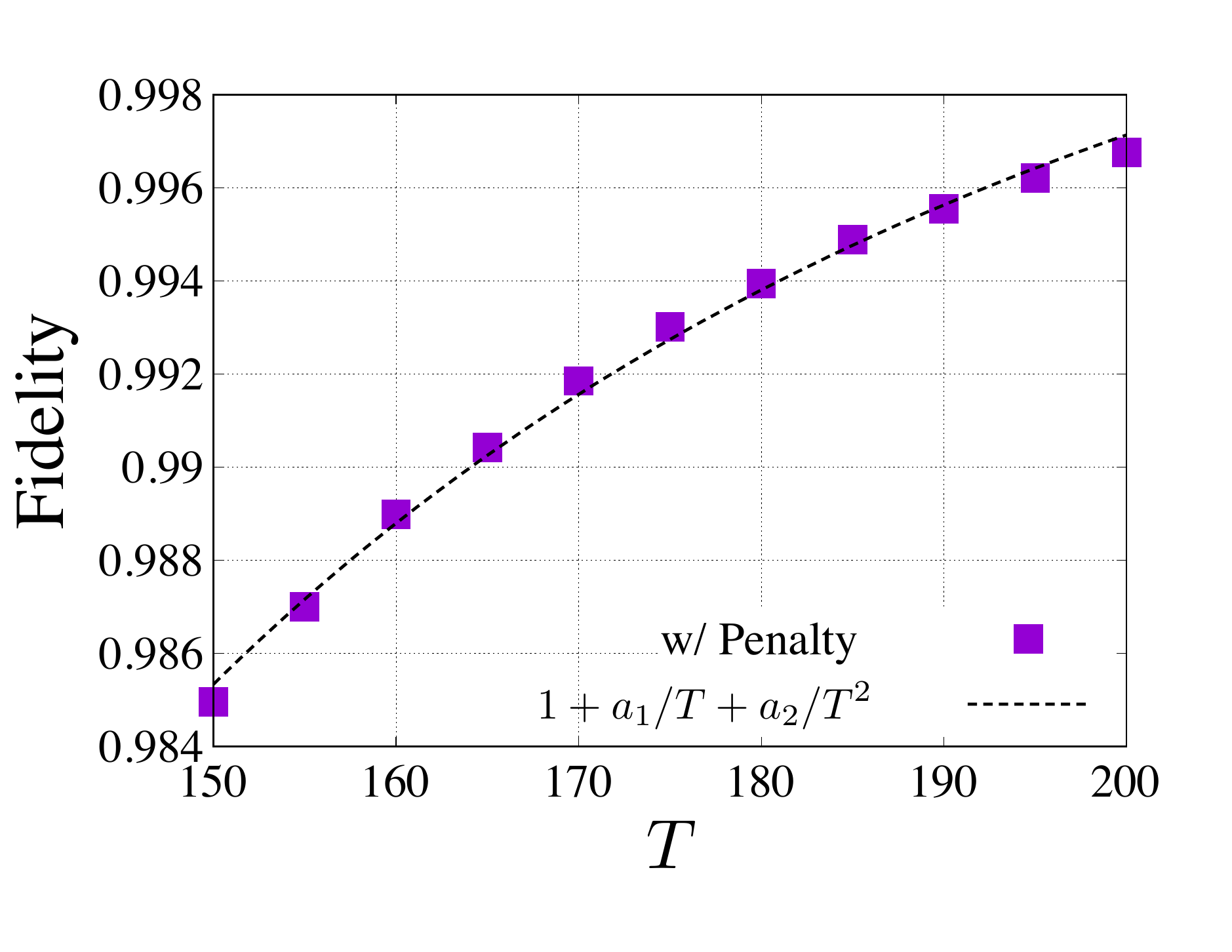}
\caption{Fidelity for the adiabatic Grover search with penalty term as functions of $T$. The system size is $L=10$.
}
\label{fig8}
\end{figure}

\section{Comparison of penalty term with non-stoquastic Hamiltonian}
In this appendix, we compare the transition matrix with the penalty term and the non-stoquastic Hamiltonian in the $p$-spin model. 
The following anti-ferromagnetic interaction is introduced as the non-stoquastic Hamiltonian \cite{Seki_2012}:
\begin{align}
\label{Nonsto}
\hat{H}_{\rm NS}=L\left(\frac{1}{L}\sum_{i=1}^{L}\hat{\sigma}^{x}_{i}\right)^2.
\end{align}
We use a QA Hamiltonian with the non-stoquastic Hamiltonian given as follows:
\begin{align}
     \hat{H}(t)=\frac{t}{T}\{\lambda \hat{H}_{p}+(1-\lambda)\hat{H}_{\rm NS}\}+\left(1-\frac{t}{T}\right) \hat{H}_{d},
\end{align}
where $\hat{H}_{p}$ is given by Eq.~(28) and $\hat{H}_{d}=-\sum_{i=1}^{L}\hat{\sigma}^{x}_{i}$. To avoid a first-order phase transition, we set $p=5$ and $\lambda=0.1$ \cite{Seki_2012}.
In Fig.~9, we show the scaling of the transition matrix in cases with and without our penalty term, and with the non-stoquastic Hamiltonian.
It can be observed that the non-stoquastic Hamiltonian does not lead to an exponential increase in the size of the transition matrix, even with an exponentially wide gap at the first-order phase-transition point. According to our general framework, opening the gap using the penalty term leads to an exponential increase in the size of the transition matrix. In contrast to our penalty term, the non-stoquastic Hamiltonian (\ref{Nonsto}) changes not only the eigenenergy but also the energy eigenstate.
\begin{figure}[!t]
\includegraphics[scale=0.4,bb=580 30 0 450]{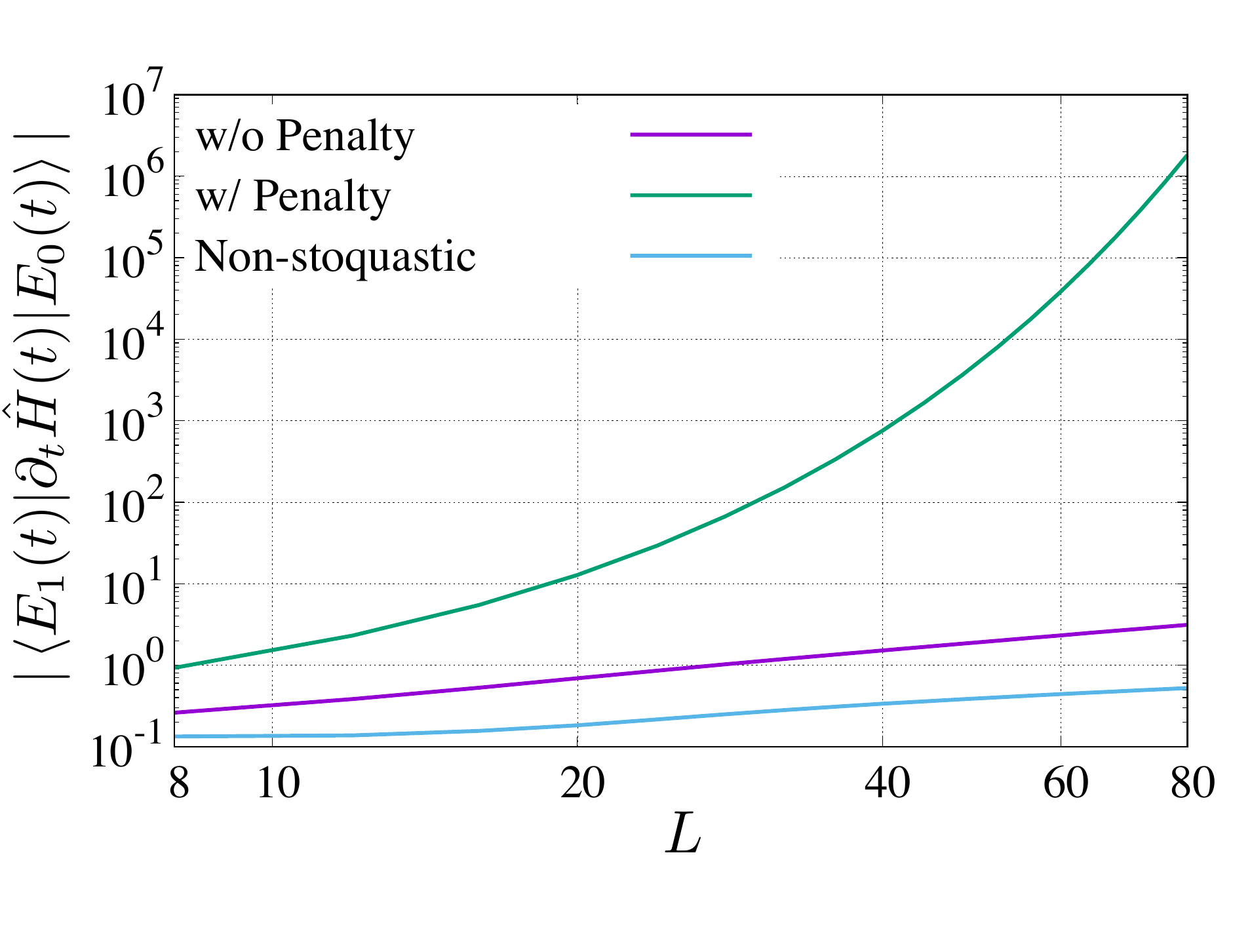}
\caption{Maximum of transition matrix for the ferromagnetic $p$-spin model. 
Purple (middle), green (top), and blue (bottom) lines denote cases without and with penalty term and with non-stoquastic term, respectively. Annealing time $T=20$.
}
\label{fig9}
\end{figure}
\bibliography{apsbib}
\end{document}